# Force-dependent bond dissociation explains the rate-dependent fracture of vitrimers

Zhaoqiang Song,‡ Tong Shen,‡ Franck J. Vernerey,* Shengqiang Cai*

**ABSTRACT:** We investigate the rate-dependent fracture of vitrimers by conducting a tear test. Based on the relationship between the fracture energy and the thickness of vitrimer films, we, for the first time, obtain the intrinsic fracture energy and bulk dissipation of vitrimers during crack extension. The intrinsic fracture energy strongly depends on tear speed, and such dependence can be well explained by Eyring theory. In contrast, the bulk dissipation only weakly depends on tear speed, which is drastically different from observations on traditional viscoelastic polymers. We ascribe such a weak rate-dependence to the strong force-sensitivity of the exchange reaction of the dynamic covalent bond in the vitrimer.

Vitrimers have recently emerged as a promising class of polymer for a variety of potential applications.[1] Due to the presence of associative exchange reactions in the polymer network, vitrimers combine desired features of both thermosets and thermoplastics, such as stable high-temperature properties together with distinct reprocessability. Although intensive research has been conducted for understanding the constitutive properties of these new materials, their fracture behaviors have been largely unexplored.

Continuous exchange reactions of dynamic covalent bonds make vitrimers "strong" viscoelastic materials[2], namely, the dependence of their viscosity on temperature can be described by Arrhenius law. Despite intense efforts made in the past, quantitative modeling or prediction of fracture in viscoelastic polymers remains challenging.[3] The difficulties are mainly two-fold: first, the molecular origin of the fracture process, together with its dependence on rate is still unclear; this both blurs the interpretation of experimental results and hinders the rational design of tough polymers;[4] second, during crack propagation, bulk dissipation often competes with the crack propagation process in a complex and rate-dependent manner[5]. To tackle the first challenge, either simple theoretical models such as the Lake-Thomas theory or empirical cohesive laws have been adopted.[4] The issue with these models, however, is that they are usually based on the assumption that the intrinsic fracture energy or the cohesive law is rate independent. To overcome the second difficulty, a linear rheology model is often adopted for viscoelastic polymers when modelling its fracture.[3,6-10] As a result, the bulk dissipation in the material caused by crack propagation is often quite rate-sensitive.[3,9] However, those simplifications are often not sufficient to accurately capture the complex fracture phenomenon observed in viscoelastic polymers.[11]

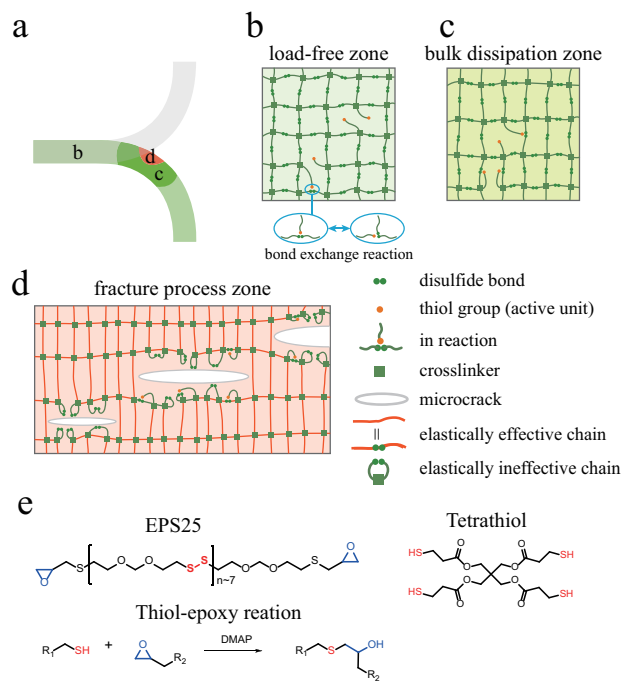

**Figure 1.** (a) Fracture process zone and bulk dissipation zone in the vitrimer in one of two arms of the specimen during the tearing test from side view. (b) Associative exchange reactions (thiol-disulfide exchange reactions in Figure S1) can happen in load-free zone. The thiol group is from the crosslinking molecules. (c) Associative bond exchange reactions occur more frequently in the bulk dissipation zone. (d) The formation of microcracks in fracture process zone reduces the areal density of elastically effective crosslinks ($\Sigma_b$) and increases the density of elastically ineffective crosslinks. (e) Chemical structures of the monomer (EPS25) and the crosslinker (tetrathiol) for synthesizing the vitrimer in this study; crosslinking is achieved through thiol-epoxy reaction with DMAP (4-Dimethylaminopyridine) as the catalyst.

In this letter, we aim to reveal critical insights into the fracture process of viscoelastic vitrimers through a combined experimental study and theoretical analysis. For this, we first conduct tear experiments on thin vitrimer films. By varying the thickness of the film, we can experimentally measure the dissipated energy in the fracture process zone (also known as intrinsic fracture energy) as well as the bulk dissipation in the film accompanied with crack extension (Figure 1). We find that the intrinsic fracture energy is highly rate-dependent, which is in contrast with the (rate-independent) assumption adopted in most previous studies on the fracture of viscoelastic polymers.[3,9] Using concepts from the classical Eyring theory, we are then able to successfully explain the scaling relationship between the vitrimer's intrinsic fracture energy and the tearing speed. This study also reveals that the bulk dissipation in the vitrimer film is only weakly dependent on the tearing rate within a range that spans three orders of magnitude; this finding is in contrast to previous studies on conventional viscoelastic polymers and gels[12]. We ascribe such weak rate-dependence of the bulk dissipation to the strong force sensitivity of the dynamic exchange reaction. Taken together, our study provides the first demonstration that thin-film tear test can be used to reveal critical insights into the fracture behavior of vitrimers.

Figure 2a shows the schematic of the tear test adopted in our study, where a vitrimer film, glued to an inextensible backing layer, is introduced with a cut along its center line. More details of the material can be found in Supporting Information (SI) and references[13,14]. The two separated arms are then stretched vertically in the opposite directions by a constant relative speed $v_0 = 2v$. Since the backing layers constrain the sample deformation, the crack is forced to propagate accordingly at a velocity $v$. Compared to many other fracture tests such as pure shear, single-edge notched tension, compact tension and center cracked tension, tear tests adopted in our experiment, similar to peeling test for measuring adhesion, have two advantages. First, the crack propagation speed can be easily controlled by the pulling speed. Second, by changing the sample thickness, the intrinsic fracture energy and the bulk dissipation can be simultaneously measured based on a simple scaling law without needing specific material models[15,16].

In the experiments, we measure the force-extension relation of thin vitrimer films with different thicknesses ($h$ =1 mm, 1.5 mm, 2 mm and 2.5 mm) and different tearing speeds ($v_0/h$=0.01/s, 0.1/s, 1/s and 10/s) as shown in Figure 2c and Figure S2. To analyze our experiment results, we normalize the crack velocity by introducing the normalized crack velocity $\bar{V} = v_0\tau/h$, where $\tau$=42 s is the characteristic relaxation time of the vitrimer determined by its stress relaxation measurement at a small strain (5%) as shown in Figure S3(a), and $h$ is the film thickness as shown in Figure 2a. It is noted that the characteristic relaxation time of the vitrimer is very close to the time scale obtained from creep tests (Figure S3(b)) and the small-amplitude oscillation shear test (Figure S3(c)). In the experiment, the normalized crack velocity $\bar{V}$ is varied between 0.42 to 420. When $\bar{V}$ is larger than 1, the crack surface is smooth as shown in Figure 2b, indicating quasi-brittle fracture. In each experiment, the pulling force oscillates around a plateau value ($F$) after an initial increase (Figure 2c), indicating the stick-slip crack growth[16]. Here we do not study the stick-slip dynamics, and simply use the average tear force on plateau $F$ to estimate the energy release rate $G$. Because of the simple geometry and the inextensible backing layer, the energy release rate can be given by $G=2F/h$.[16]

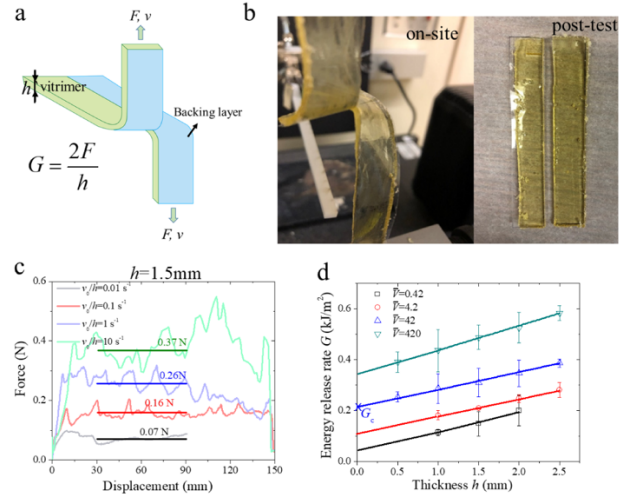

**Figure 2.** (a) A schematic of tear test. (b) Tearing the vitrimer and the torn vitrimer with a smooth crack surface. (c) Force vs. Displacement of tear experiments with the film thickness of 1.5mm. (d) Energy release rate as a function of thickness of the films with different normalized crack velocities $\bar{V}$.

In Figure 2d, we plot the energy release rate $G$ as a function of the film thickness $h$ for four different crack velocities $\bar{V}$. We find that the relationship between the energy release rate and the film thickness is linear for all of the four different crack velocities ($\bar{V}$). Such linear relationship indicates that the size of the fracture process zone or "fractocohesive length" is much smaller than the film thickness[15,16], as shown in Figure 1a, so we can separate the fracture energy into two parts:

$$G = G_c + G_d, \qquad (1)$$

where $G_c$ is the energy dissipation (intrinsic fracture energy) in the fracture process zone (Figure 1d), and $G_d$ is the energy dissipated by viscoelastic loss in the bulk of the film (Figure 1c).

Based on de Gennes's model for fracture in a viscoelastic materials,[17] the size of bulk dissipation zone in the material is given by $R_d = \frac{E_0}{E_\infty}R_0$, where $R_0$ is the fractocohesive length, $E_0$ is the instant modulus, and $E_\infty$ is the equilibrium modulus. For a vitrimer, the equilibrium modulus is zero, so the characteristic size of bulk dissipation in a vitrimer is infinitely large. As a result, the film thickness $h$ is the only relevant length scale for bulk dissipation and thus $G_d \propto h$.[15,17] Our experimental data in Figure 2d also suggests the linear dependency of $G_d$ on $h$, so $G = G_c + w_p h$, where $G_c$ is independent of $h$ and $w_p$ is the average density of bulk dissipation.[18] Based on this equation, the intersection of the fitted linear relationship between $G$ and $h$ with the vertical axis gives the intrinsic fracture energy $G_c$, as shown in Figure 2d. By measuring the slopes of the fitting lines relating the energy release rate $G$ and the thickness $h$ in Figure 2d, we can further measure the magnitude of $w_p$.

We envision different energy dissipation processes in the area near the crack tip and in the bulk. In the bulk, the energy dissipation is mainly through viscoelasticity resulted from dynamic exchange reactions. However, moving towards the crack tip, the chains are increasingly stretched due to the stress concentration. According to Eyring's theory[19], this leads to an increase in the bond exchanging rate, facilitating the formation of cavities or microcracks. With associative bond exchange, though the crosslink density remains unchanged in the vitrimer, the elastically effective chains may become elastically ineffective (such as loops as shown in Figure 1d). Therefore, within a small region near the crack tip, the network loses its integrity as damage occurs.

Figure 3a plots the intrinsic fracture energy $G_c$ as a function of the normalized crack velocity ($\bar{V}$). The strong rate-dependent intrinsic fracture energy of our vitrimer is in contrast to the assumption of rate-independent intrinsic fracture energy of viscoelastic polymers commonly adopted in previous studies[3,9,12]. It is also noted that the magnitude of $G_c$ is much larger than the value predicted from the Lake-Thomas model (see Supporting Information). As shown in Figure 1, we believe that a fracture process zone, which is much larger than the mesh size of the polymer network, exists near the crack tip during tearing. To explain the scaling relationship between $G_c$ and the crack velocity ($\bar{V}$), we extend the theory developed by Chaudhury and Hui[19,20] and assume that in the fracture process zone, the energy dissipation is mainly caused by accelerated bond dissociation. Similar to the picture depicted by Lake and Thomas[21], the energy stored (denoted as $W$) in a polymer chain is entirely dissipated and we have $G_c \sim \Sigma W$, where $\Sigma$ is areal density of polymer chains.

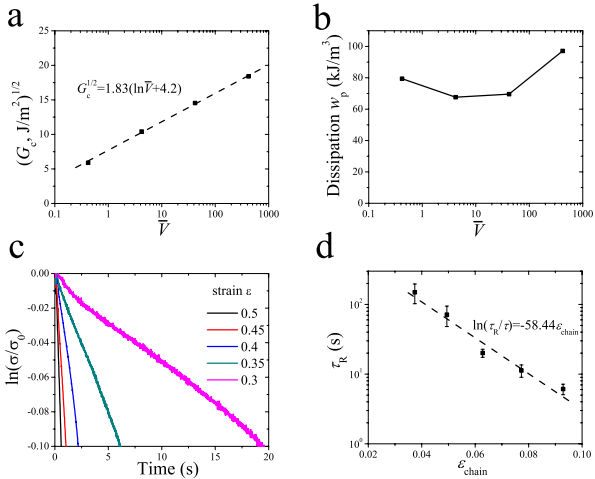

**Figure 3.** (a) The intrinsic fracture energy as a function of the normalized crack velocity, where the intrinsic fracture energies are the interceptions of the lines with vertical axis in Figure 2d; (b) Bulk dissipation of vitrimer with different normalized crack velocities, where bulk dissipation is the slopes of the lines in Figure 2d. (c) Stress relaxation of a vitrimer with different applied strains $\varepsilon$. (d) Relaxation time as a function of the chain strain $\varepsilon_{chain}$.

According to Eyring theory, the force $f$ applied to a chain modifies the energy landscape for bond dynamics. This leads to an increase of bond exchanging rate $k_d = \tau^{-1} \exp\left(\frac{f\Delta_a}{k_BT}\right)$, where $\tau$ is the natural frequencies for bond exchanging[22]. In the fracture process zone, we assume that polymer chains that experience exchange reaction become elastically ineffective and lose their load-bearing capacity (Figure 1d). Consequently, the decrease in the areal density $\Sigma_b$ of elastically effective crosslinks follows the rate equation:

$$-\frac{D\Sigma_b}{Dt} = \tau^{-1} \exp\left(\frac{f\Delta_a}{k_BT}\right)\Sigma_b, \qquad (2)$$

where $k_B$ is the Boltzmann constant, $\Delta_a$ is the activation length of the covalent bond, $T$ is the absolute temperature. The frequency of bond exchanging follows the Arrhenius law, $\tau = k_0 \exp(\frac{E_a}{k_BT})$, in which $k_0$ is a precursor and $E_a$ is the activation energy of bond exchanging without external force. Considering linear chains with spring stiffness $k_s$, the force is related to deformation as $f = k_s\delta = k_s\varepsilon_{chain}L_0$, where $\delta$, $\varepsilon_{chain}$ and $L_0 = \sqrt{n}l$ are respectively the extension, strain and end-to-end distance of the polymer chain in a free standing state with $n$ and $l$ the number and length of Kuhn segments in a chain, respectively.

The force-sensitivity of bonds described in Eq. (2) can be experimentally explored by subjecting the vitrimer to a stress relaxation test. Indeed, during the relaxation process, once a polymer chain dissociates at fixed strain it no longer contributes to the network mechanics since free chains reassociate in a stress-free state, and the chain reassociation does not contribute to the stress. Therefore, the characteristic relaxation time of the network when a force $f$ is applied on a polymer chain in Eq. (2) is $\tau_R = \tau \exp\left(-\frac{f\Delta_a}{k_BT}\right)$. With the linear spring assumption of the polymer chain, we find that:

$$\ln\left(\frac{\tau_R}{\tau}\right) = -\frac{k_sL_0\Delta_a}{k_BT}\varepsilon_{chain}. \qquad (3)$$

This implies that the force sensitivity ($\frac{k_sL_0\Delta_a}{k_BT}$) of the bond is a material parameter that remains constant during stress relaxation. The experimental results of the relaxation of vitrimer at different strains are shown in Figure 3c. We find that the stress reduction follows an exponential decay as: $\sigma/\sigma_0 = \exp(-t/\tau_R)$. To extract the force sensitivity of dynamic covalent bond, we employ the eight-chain model[23] and extract the strain of the polymer chain as $\varepsilon_{chain} = \sqrt{I_1/3} - 1$, where $I_1$ is the first invariant of the left Cauchy-Green deformation tensor and $I_1 = (1+\varepsilon)^2 + 2/(1+\varepsilon)$ for uniaxial tension with $\varepsilon$ the tensile strain. Figure 3d shows the experimental result of relaxation time as a function of the strain, where a linear relationship between $\ln(\tau_R/\tau)$ and $\varepsilon_{chain}$ is found and can be fitted by Eq. (3). The slope of line can be used to extract the force-sensitivity of dynamic disulfide bonds as $k_sL_0\Delta_a/(k_BT) = 58.44$. This force sensitivity can be also estimated as follows. The spring stiffness of a polymer chain can be estimated by the modulus of the elastomer $E$ and the mesh size of the network $L_0$ as $k_s \sim EL_0$, where $E \sim 1$ MPa is Young's modulus[13] and $L_0 \sim 10$ nm (see SI) of our vitrimer. So, the spring stiffness of polymer chain $k_s$ is in an order of 0.01 N/m. With the consideration of the activation length of disulfide bonds $\Delta_a \sim 0.3$ nm and temperature $T \sim 300$ K, we can estimate the force sensitivity is $\frac{k_s L_0 \Delta_a}{k_B T} \sim 10$, which has the same order of the value of force sensitivity measured from stress relaxation in figure 3d.

We next investigate how the role of the bond's force sensitivity on the rate-dependent intrinsic fracture energy measured in the experiment (Figure 3a). We rewrite Eq. (2) as

$$v_c \frac{d\Sigma_b}{d\delta} = -\frac{1}{\tau}\Sigma_b \exp\left(\frac{k_s\delta\Delta_a}{k_BT}\right), \qquad (4)$$

where the velocity of the chain stretch is $v_c = d\delta/dt$. $v_c$ can be estimated from the crack velocity as follows: due to the existence of a backing layer, the material deformation is constrained into a small region of length $h$ around the moving crack tip[16]. Therefore, a characteristic time for chain deformation is obtained as $t_0 = h/v_0$. Furthermore, since the surface is fully separated at the end of the deformation zone, we assume that chains are elongated from their natural length $\sqrt{n}l$ to the contour length $nl$ when they travel through the deformation zone. Based on this conceptual picture, the velocity of the chain stretch can be estimated as $v_c = (\sqrt{n} - 1)L_0 v_0/h$ in Eq. (4). An average dissociation length of polymer chain can be defined as $\bar{\delta} = \int_0^\infty \frac{\Sigma_b}{\Sigma_0} d\delta$, where $\Sigma_0$ is total number of chains per unit area. By recalling Eq. (4) and following the derivation of Chaudhury[19], the integration can be obtained as $\bar{\delta} = \frac{k_B T}{k_s \Delta_a} \ln\left(\frac{(\sqrt{n}-1)k_s \Delta_a L_0}{k_B T}\bar{V}\right)$, if $k_s \Delta_a L_0 \bar{V} \gg k_B T$.[20, 24] The average force on a chain before failure can be expressed as $f_{break} = k_s \bar{\delta}$ and the average energy stored in a polymer chain before breakage is $W = \frac{1}{2} k_s \bar{\delta}^2$. Invoking our scaling analysis, the intrinsic fracture energy $G_c$ therefore scales as:

$$G_C \propto \left[\ln\left(\frac{(\sqrt{n}-1)k_s \Delta_a L_0}{k_B T}\bar{V}\right)\right]^2, \quad (5)$$

or $\sqrt{G_c} = \alpha(\ln(\bar{V}) + \beta)$, where the coefficient $\alpha$ is related to the thermally activated force-sensitivity of dynamic covalent bonds and the size of the fracture process zone, coefficient $\beta$ is a rate-independent constant expressed as

$$\beta = \ln\left(\frac{(\sqrt{n}-1)k_s \Delta_a L_0}{k_B T}\right). \quad (6)$$

From Eq. (5), we know that $\sqrt{G_c} \propto \ln\bar{V}$, which agrees well with the result shown in Figure 3a. A good agreement between model and experiment is further obtained with $\beta = 4.2$ as shown by the fitting shown in Figure 3a.

We can also evaluate the coefficient $\beta$ defined in Eq. (6), based on that $k_s L_0 \Delta_a/(k_B T) = 58.44$, determined from the stress relaxation measurements previously, and the number of Kuhn segments $n$. To estimate the number of Kuhn segments of the vitrimer for our experiment, we conducted uniaxial tension tests on a thin vitrimer strip with a strain rate of 100%/s. For such a strain rate, the bond exchange reaction is negligible during the deformation. We found that the rupture strain of the vitrimer strip is around 150% as shown in Figure S4. According to the eight-chain model,[23] the chain locking stretch (defined as the stretch of a chain when it is fully extended) can be estimated from the rupture strain of the vitrimer strip as $\lambda_L = 1.54$, under the assumption that the polymer chain is fully extended at rupture. The number ($n$) of Kuhn segments could further be related to the chain locking stretch ($\lambda_L$) as $\lambda_L = \sqrt{n}$, which yields $n = 2.4$. Consequently, we can obtain $\beta = 3.5$, which is close to the one estimated from the measurement of the rate-sensitivity of the intrinsic fracture energy shown in Figure 3a.

Another notable result of this study pertains to the rate of bulk dissipation $w_p$. Indeed, Figure 3b suggests that it is only weakly dependent of the tearing rate, varying between 65~98 $kJ/m^3$ over the range of $\bar{V}$ from 0.42 to 420. This weak rate-dependence is in sharp contrast to the rate-dependent bulk dissipation of most viscoelastic polymers studied in the past[12]. For a linear viscoelastic material, it is expected that bulk dissipation should reach a maximum when the loading time is comparable to the single relaxation time of the material, and reach a minimum when the loading rate is too small or too large, as shown in Figure S5(a-b). We postulate that such difference is caused by the nonlinear viscoelasticity of the vitrimer, where the rate of bond exchange increases at larger forces or strains ($\bar{V}$) and thus there is no single relaxation time in the material. The stretch-stress curve of a strongly force-sensitive vitrimer (using the parameter $k_s L_0 \Delta_a/(k_B T) = 58.44$) experiencing loading-unloading history at different loading rates, is calculated by combining the classic transient network theory with Eyring's theory and steady-state kinetics for chain dissociation and reassociation,[5] as shown in Figure S5c. The strong force-sensitivity of dynamic covalent bonds leads to large bulk dissipation even when $\bar{V}$ is very large e.g. $\bar{V}$ =42, 420, as shown in Figure S5d. A more detailed study of $w_p$ in the tear test of a vitrimer film requires a full-field simulation of the tearing process with a nonlinear viscoelastic material model, which is a challenging task and has not been achieved in the literature yet. However, in this work, we circumvent this difficulty and simply extract the effect of $w_p$ on the fracture energy based on geometric scaling.

In summary, our study clearly shows that tear test of a vitrimer film is an effective way to explicitly reveal the molecular origins of rate-dependent energy dissipation associated with its fracture. Through the experiment, we find that the intrinsic fracture energy of vitrimers with disulfide bonds is highly rate-dependent while their bulk dissipation is rather rate-insensitive during fracture; this is significant contrast to most previously studied viscoelastic polymers. By assuming a small fracture process zone near the crack tip within which the acceleration of bond dissociation is the main energy dissipation mechanism, we could explain the scaling relationship between the intrinsic fracture energy and the normalized crack velocity based on the classical Eyring theory.

## ASSOCIATED CONTENT

**Supporting Information**. Material preparations and Mechanical characterizations and additional calculations.
This material is available free of charge via the Internet at http://pubs.acs.org."

## AUTHOR INFORMATION

**Corresponding Authors**

**Shengqiang Cai** - Department of Mechanical and Aerospace Engineering, and Materials Science and Engineering Program, University of California, San Diego, La Jolla, CA 92093, USA; Email: shqcai@ucsd.edu

**Franck J. Vernerey** - Department of Mechanical Engineering, and Program of Materials Science and Engineering, University of Colorado Boulder, Boulder, CO, 80302, USA; Email: franck.vernerey@colorado.edu

**Authors**

**Zhaoqiang Song -** Department of Mechanical and Aerospace Engineering, University of California, San Diego, La Jolla, CA 92093, USA

**Tong Shen -** Department of Mechanical Engineering, University of Colorado Boulder, Boulder, CO, 80302, USA

**Author Contributions**

‡These authors contributed equally.

**Notes**

The authors declare no competing financial interest.

## ACKNOWLEDGMENT

S.C acknowledges financial support from the Office of Naval Research through the Grant ONR N00014-17-1-2056. F.J.V. acknowledges financial support from the National Science Foundation under award no.1761918.

# Supporting Information for

# **Force-dependent bond dissociation explains the rate-dependent fracture of vitrimers**

Zhaoqiang Song, Tong Shen, Vernerey Franck*, and Shengqiang Cai*

Corresponding to: V. Franck, email: franck.vernerey@colorado.edu; S. Cai, email: shqcai@ucsd.edu

1. Material synthesis

EPS25 (epoxy equivalent = 777 g/equiv) was kindly provided by Akzo Nobel Chemicals; pentaerythritol tetrakis(3-mercaptopropionate) (PETMP, Sigma Aldrich), and 4-(dimethylamino)pyridine (DMAP, Sigma Aldrich) were used as received.

We synthesized EPS25 vitrimer by following the previous works.[1,2] For the synthesis of vitrimer, we added 5 mmol PETMP into 10 mmol EPS25, and stirred the mixture until homogeneous. Then, we added 1 wt% of DMAP as catalyst into the mixture. We stirred the mixture for 10 minutes and degassed under vacuum. After that, we poured the mixture into a glass mold and heated at 60°C for 2 hours. Finally, we obtained the vitrimer. The epoxy group in EPS25 reacted with the thiol group in PETMP under the catalyst DMAP, as shown in Figure 1e. We conduct all the experiments on the vitrimers within one day after its synthesis.

2. Mechanical characterization

    2.1 Stress relaxation test

We conducted the stress relaxation experiments by using a tensile testing machine (Instron 6950, Norwood) with a 10 N load cell. In the stress relaxation experiments, we applied instant strains of 5%, 30%, 35%, 40%, 45% and 50% for the samples and measured the stress as a function of time. We repeated each test for three times.

    2.2 Creep test and small-amplitude oscillation shear (SAOS) test

We conducted the creep test and SAOS test by using the Discovery HR-3 Rheometer (TA Instruments). All tests are conducted with a 20-mm steel Peltier plate and 1.1-mm gap size. In the creep test, we applied a time-independent shear stress of 100 Pa onto the sample and measured the strain as a function of time. We repeated the test for three times. The SAOS test is conducted with a fixed temperature (23°C).

2.3 Tear test

The vitrimer samples were prepared with 30 mm in width and 90 mm in length. We cut each inextensible backing layer (3M Scotch Sure Start Packaging tape, with a thickness of 0.066 mm) into a long rectangular shape (with a width of 15 mm and a length of 90 mm), and glued it on half of each surface of the vitrimer sample, without any observable gap between the long edges of the backing layers, as shown in Figure 2a. Then we introduced a pre-crack along the center line of the vitrimer with the length of 15 mm by using a sharp blazer, forming two arms of 15 mm width and 15 mm length. The two arms of the sample were then fixed by the grips of a tensile testing machine (Instron 6950) with a 10 N load cell. During the tear, the tensile testing machine applied a constant loading speed and recorded the force (Figures 2c and Figure S2). We repeated each test for three times.

3. Intrinsic fracture energy of Lake-Thomas model

The Lake–Thomas model[3] assumes that the intrinsic fracture energy is the chemical energy stored in the unit area of a single-chain layer:

$$\Gamma_0 = \frac{l\sqrt{\bar{n}}U}{V}, \quad (S1)$$

where $l$ is the length of the monomer, $\bar{n}$ the average number of monomers in a chain, $U$ is the chemical energy of the bonds (C–C bonds, C-O bonds, C-S bonds and S-S bonds) in the backbone of a monomer, and $V$ is the volume of a monomer. For a EPS25 monomer, the number of C–C bonds in per monomer is $n_0(C-C) = 21$, the number of C–O bonds in per monomer is $n_0(C-O) = 32$, the number of S–S bonds in per monomer is $n_0(S-S)=7$, and the number of C–S bonds in per monomer is $n_0(C-S)=16$. The C–C bond energy is $U_{C-C}=1.378\times10^{-19}$ J, the C–O bond energy is $U_{C-O}=1.395\times10^{-19}$ J, the S-S bond energy is $U_{S-S}=0.847\times10^{-19}$ J, and the C–S bond energy is $U_{C-S}=1.0299\times10^{-19}$ J in average.[4] Taking into account of the number of bonds and the bond energy in a monomer, the chemical energy $U=1.14\times10^{-17}$ J. The volume of a monomer can be estimated as

$$V = \frac{M}{N_A \rho} = 2.15\times10^{-27} \text{ m}^3, \tag{S2}$$

where $N_A$ is the Avogadro number ($6.022\times10^{23}$), $\rho$ is the density of EPS25 (1.2 g/cm³), $M$ is the molecular weight of EPS25 (1554 g/mol). The length of monomer is $l=\sqrt[3]{V}=1.29\times10^{-9}$ m. The average number of monomers in a chain is

$$\bar{n} = \frac{1}{VN} = 9.6, \tag{S3}$$

where $N = \mu/k_B T = 4.826\times10^{25}$ m$^{-3}$ is the number of polymer chains per unit volume with $\mu$ shear modulus of EPS25-resin epoxy, $k_B$ Boltzmann constant (1.3806×10$^{-23}$ m² kg s$^{-2}$ K$^{-1}$) and $T$ the temperature. Thus, the intrinsic fracture energy of the vitrimer is $\Gamma_0 = 21.2$ J/m². The natural length of polymer chain is $L_0 = \bar{n}l/\sqrt{n}$=8 nm with the number of Kuhn segments $n = 2.4$.

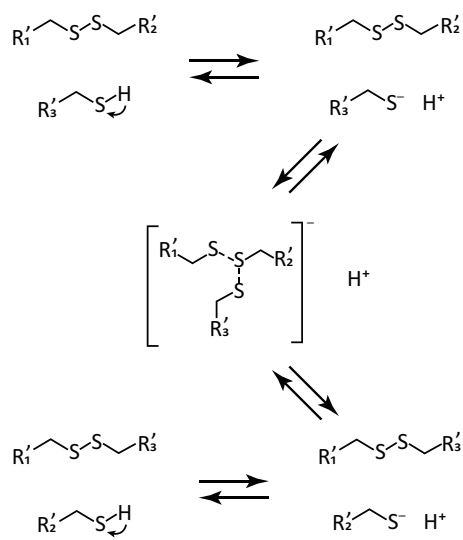

Figure S1. In EPS25 vitrimer, thiol-disulfide exchange reaction is divided into three steps: deprotonation of the thiol group, nucleophilic attack of the thiolate on one of the sulfur atoms of the disulfide, and dissociation of the intermediate product.[5]

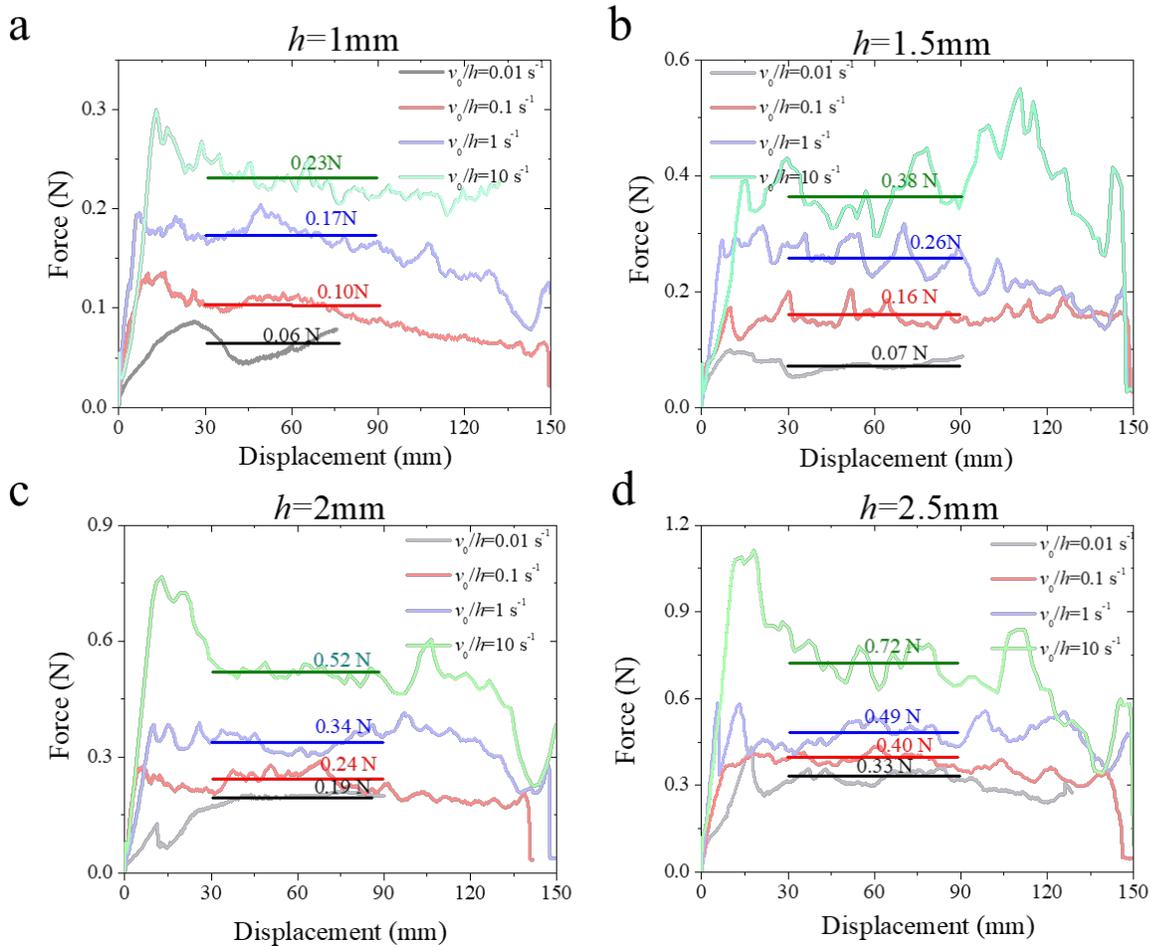

Figure S2. Force-displacement of tear experiments with different thickness (a) 1mm, (b) 1.5mm, (c) 2mm, (d) 2.5mm.

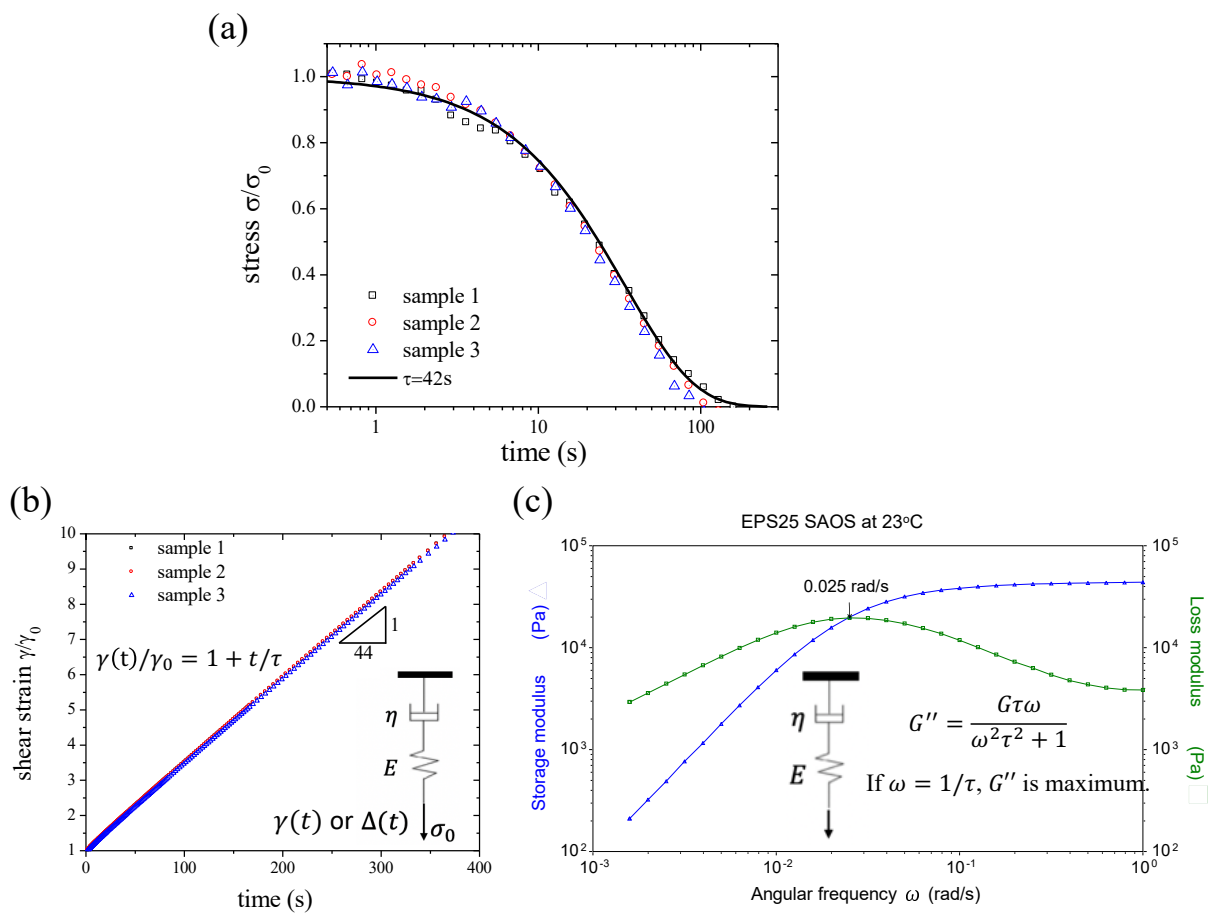

Figure S3. (a) Stress relaxation of vitrimer at room temperature (23°C) with a small applied strain (5%), and the characteristic relaxation time predicted from Maxwell model is 42 s. (b) Creep of EPS25 at 23°C with the characteristic time as 44 s; (c) The small-amplitude oscillation shear experiment with the terminal relaxation time 40 s.

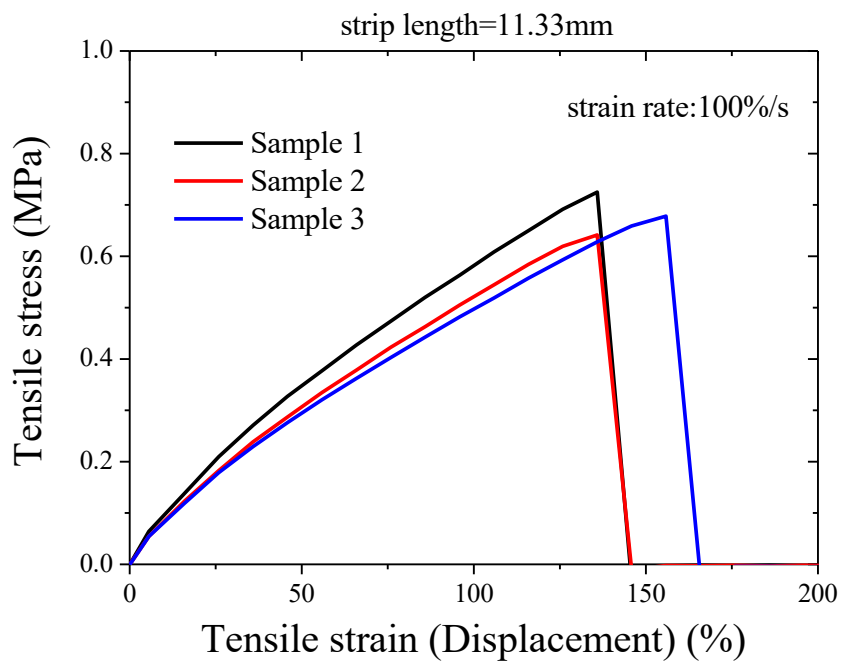

Figure S4. The uniaxial tension test of vitrimer strips. The strain rate is 100%/s and the strip length is 11.33 mm.

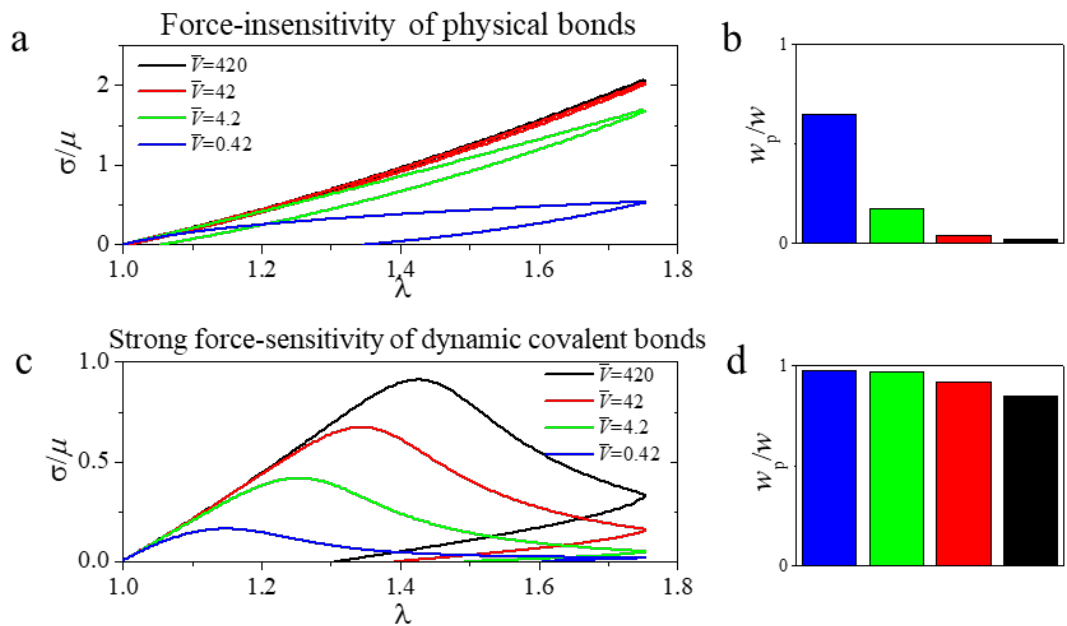

Figure S5. (a) Stress-strain curve of a force-insensitivity polymer experiencing loading-unloading history at different loading rates ($\bar{V}$). (b) The proportion of dissipation energy ($w_p$) in total strain energy ($w$) of a force-insensitivity polymer. (c) Stress-strain curve of a strong force-sensitivity vitrimer experiencing loading-unloading history at different loading rates. (d) The proportion of dissipation energy ($w_p$) in total strain energy ($w$) of a strong force-sensitivity vitrimer.

14